\documentclass[notitlepage,a4paper,aps,prd,superscriptaddress,nofootinbib,groupedaddress]{revtex4}

\usepackage{enumerate}
\usepackage{amsmath}
\usepackage{amsfonts}
\usepackage{amssymb}
\usepackage[utf8]{inputenc}
\usepackage[T1]{fontenc}
\usepackage{mathtools}
\usepackage{wasysym}
\usepackage{accents}
\usepackage{graphicx}
\usepackage[english]{babel}
\usepackage[svgnames]{xcolor}
\usepackage[colorlinks=true,citecolor=blue]{hyperref}
\usepackage{url}
\usepackage{amsmath}
\usepackage{float}
\usepackage{mhchem}
\usepackage{bbold}
\usepackage{subcaption}

\newcommand{\JP}{
\affiliation{Physics Department, Federal University of Para\'iba, Caixa Postal 5008, 58059-900, Jo\~ao Pessoa-PB, Brazil}
}
\newcommand{\CG}{
\affiliation{Physics Department, Federal University of Campina Grande, Caixa Postal 10071, 58429-900, Campina Grande-PB, Brazil}
}

\newcommand{\Napoli}{
\affiliation{Dipartimento di Fisica “Ettore Pancini”, Università di Napoli “Federico II”, Napoli, Italy} 
}

\newcommand{\INFN}{
\affiliation{Istituto Nazionale di Fisica Nucleare, Sezione di Napoli, Complesso Univ. Monte S. Angelo, I-80126 Napoli, Italy} 
}

\newcommand{\Lavras}{
\affiliation{Department of Physics, Federal University of Lavras, Caixa Postal 3037, 37200-900 Lavras-MG, Brazil}
}

\begin{document}

\title{Coherence and Entanglement in a Non-commutative Spacetime}

\author{Iarley P. Lobo}
\email{lobofisica@gmail.com,\\  iarley.lobo@academico.ufpb.br}
\affiliation{Department of Chemistry and Physics, Federal University of Para\'iba, Rodovia BR 079 - km 12, 58397-000 Areia-PB,  Brazil}
\CG

\author{Gislaine Varão}\email{gislainevarao@gmail.com}
\JP

\author{Giulia Gubitosi}\email{giulia.gubitosi@unina.it}
\Napoli
\INFN

\author{Moises Rojas}\email{moises.leyva@ufla.br}
\Lavras

\author{Valdir B. Bezerra}\email{valdir@fisica.ufpb.br}
\JP

\begin{abstract}
We investigate the emergence of quantum coherence and quantum correlations in a two-particle system with deformed symmetries arising from the quantum nature of spacetime. We demonstrate that the deformation of energy-momentum composition induces a momentum-dependent interaction that counteracts the decoherence effects described by the Lindblad equation in quantum spacetime. This interplay leads to the formation of coherence, entanglement and other correlations, which we quantify using concurrence, the $l_1$-norm of coherence, quantum mutual information and Local Quantum Fisher Information. Our analysis reveals that while the openness of quantum spacetime ultimately degrades entanglement, it also facilitates the creation and preservation of both classical and quantum correlations.
\end{abstract}

\maketitle


\section{Introduction}

Over the past few decades, significant theoretical efforts have been devoted to formulating a quantum theory of gravity. Various proposals have emerged, including string theory \cite{Polchinski:1998rq}, loop quantum gravity \cite{Ashtekar:2021kfp}, causal dynamical triangulation \cite{Loll:2019rdj}, and asymptotically safe gravity \cite{Eichhorn:2018yfc}, among others. Quantum gravity proposals suggest that, at a fundamental level, the mathematical description of spacetime should incorporate Planck scale corrections, and the existence of a fundamental length has been raised as trace of such kind of departure \cite{Hossenfelder:2012jw}. A way to formalize this physical effect is encoded in the concept of a non-commutative spacetime, since at a quantum level it signals to a fundamental limit in the localization of events, that complements the Heisenberg uncertainty principle. In this paper, we shall use the term {\it quantum spacetime} to refer to this kind of structure.

One of the most intriguing aspects of these approaches is the possibility that Lorentz's symmetry may not be a fundamental symmetry of nature. At the Planck scale, this could imply either the absence of local equivalent frames or that Lorentz symmetry is an intermediate step toward a more fundamental symmetry, which is a scenario that has been explored in astrophysics \cite{Addazi:2021xuf} and tabletop experiments like concerning optical clocks \cite{Hohensee:2013cya,Pruttivarasin:2014pja,Sanner:2018atx} and quantum optics \cite{Pikovski:2011zk}. The latter scenario suggests that the principle of relativity could be compatible with the existence of a fundamental length scale \cite{Amelino-Camelia:2000stu}, a possibility currently under experimental investigation \cite{Amelino-Camelia:2008aez,Addazi:2021xuf,AlvesBatista:2023wqm}.

A quantum spacetime with nontrivial features can leave possibly observable imprints on quantum mechanical systems. These include violations of the equivalence principle \cite{Wagner:2023fmb} and the emergence of fractal dimensions \cite{Varao:2024eig}. Moreover, quantum spacetime may act as an open environment, therefore leading to fundamental decoherence in quantum systems \cite{Breuer:2008rh, Streltsov:2016iow}. This phenomenon can be associated with a violation of the relativity principle, where Lorentz (or Galilean) transformations cease to be symmetries \cite{Petruzziello:2020wkd}. Moreover, even when the relativity principle is preserved at the Planck scale, decoherence effects can still arise as a consequence of  a deformed time evolution operator. This produces a modification of the  evolution equation of quantum systems described by a Lindbladian operator \cite{Arzano:2014cya,Arzano:2022nlo}.  While this has been established for single-particle states, the effects on multiparticle systems have not been addressed in this context. In this work, we investigate this, focusing on two-particle systems.

Modified spacetime symmetry transformations \cite{Lukierski:1991pn,Lukierski:1993wxa}  can alter conservation laws \cite{Majid:1994cy,Amelino-Camelia:2000stu,Borowiec:2009vb}, thus giving rise to the concept of relative locality \cite{Amelino-Camelia:2011lvm,Gubitosi:2011hgc}. One intriguing possibility is the modification of energy conservation, which, at the quantum level, implies that the Hamiltonian of a two-particle system could acquire additional nontrivial contributions leading to decoherence, which do not have relevant implications in the analysis of single-particle decoherence.
In addition to decoherence effects, depending on the nature of the additional terms in the Hamiltonian, correlations between particles can emerge and potentially compete with decoherence effects. For instance, recent studies have explored the formation of entanglement due to gravitational interactions between massive states in spatial superpositions \cite{Bose:2017nin,Marletto:2017kzi}, offering a potential window into quantum gravity. Entanglement, a quintessentially quantum correlation with no classical counterpart \cite{Horodecki:2009zz}, is one such possibility. Beyond entanglement, quantum discord represents the first historical measure of quantum correlations \cite{PhysRevLett.88.017901}, defined as the difference between mutual information and classical correlations. 

Besides that, theoretical aspects of gravitational decoherence has been pushed the boundaries of our understanding of the relation between quantum phenomena and spacetime properties \cite{Blencowe:2012mp,Foo:2021exb,Moller:2023iwo,Fuentes:2010dt}, which corresponds to an interplay that has been explored in a pioneering way by series of works in the past years \cite{Fuentes-Schuller:2004iaz,Alsing:2006cj,Martin-Martinez:2010yva,Fuentes:2010dt}.

In this paper, we examine  quantum correlation through  the evolution of three key quantifiers: the {\it concurrence} as a measure of entanglement \cite{Wootters:1997id}, the $l_1${\it -norm of coherence} to track decoherence effects \cite{Baumgratz:2013ecx}, the {\it quantum mutual information} as discussed in \cite{PhysRevLett.88.017901} and the {\it local quantum Fisher information}, which measures how much useful information a subsystem carries about a parameter to be estimated, and is a fundamental tool for quantum estimation theory \cite{slaoui2019comparative,Silva:2025vmu,kim2018characterizing,Dahbi:2022uyi,Lian:2021xjw,Yurischev:2023jrd}.

The paper is organized as follows. In Section \ref{sec:def_sym}, we revisit the formalization of deformed symmetries within a Hopf algebra model and the derivation of a Lindblad equation. In Section \ref{sec:two_part}, we construct a model of two quantum particles with two energy levels, incorporating an interaction term arising from the deformed composition of energy. In Section \ref{sec:quant_corr}, we analyze quantum correlations in this model, explaining in more details the motivation and applications for each of them. 
In Section \ref{sec:diff_l}, we explore how the deformed Hamiltonian and the Lindblad term influence the interplay between these correlations. Finally, in Section \ref{sec:discussion}, we present our conclusions. Throughout the paper, we use natural units where $\hbar = c = 1$, implying that the Planck energy is the inverse of the Planck length.

\section{Deformation of Spacetime Symmetries}\label{sec:def_sym}

One of the most studied examples of non-commutative spacetimes is the so called $\kappa$-Minkowski spacetime given by the following commutation relation between coordinates
\begin{equation}
    \left[x^0,x^i\right]=i\ell x^i\, ,\qquad \left[x^i,x^j\right]=0,
\end{equation}
where Latin indices span space coordinates, and $\ell$ is a length scale, supposedly of the order of the Planck length.\footnote{It is also commonly used the notation $\kappa=\ell^{-1}$ to refer the energy scale $\kappa$, in natural units, as the one that governs quantum gravity corrections.} A generic feature of non-commutative spacetimes is that the description of free particles as plane waves is generalized to non-commutative ordered exponentials \cite{Majid:1994cy} $:e^{k_{\mu}x^{\mu}}:$, since the parameters appearing in these exponentials are coordinates of a non-Abelian Lie group. 

This means that if quantum states are given in the momentum representation, they are labeled by momentum coordinates on the group $|k(g)\rangle$ such that the action of the translation generators, which are the spatial momentum operators, is $P_{\mu}|k(g)\rangle=k_{\mu}(g)|k(g)\rangle$. As a consequence, these momentum eigenvalues obey a deformed composition law $k_{\mu}(g)\oplus k_{\mu}(h)=k_{\mu}(gh)$. From this, we can define the inverse group element $k(g^{-1})$ in which $k(g)\oplus k(g^{-1})=0$. The action of the momentum operator on the dual space elements allows one to define the antipode operator $S(P)$ such that $P_{\mu}\langle k(g)|=k_{\mu}(g^{-1})\langle k(g)|=\langle k(g)|S(P_{\mu})$. Besides that, the representation of the momentum operator on a two-particle state is given by the coproduct map
\begin{equation}
    \Delta P_{\mu}\left(|g\rangle\otimes |h\rangle\right)=k_{\mu}(gh)\left(|g\rangle\otimes |h\rangle\right)\, .
\end{equation}

This general approach finds a natural application in the $\kappa$-Minkowski spacetime (see, for instance \cite{Arzano:2014cya,Arzano:2022nlo} and references therein), in which these structures are given by
\begin{align}
S(P_0)&=-P_o+\ell {\bf P}^2\Pi_0^{-1}\, , \\
    S(P_i)&=-P_i\Pi_0^{-1}\, ,\\
    \Delta P_0&=P_0\otimes \Pi_0+\Pi_0^{-1}\otimes P_0+\ell \sum_i P_i\Pi_0^{-1}\otimes P_{i}\, ,\\
    \Delta P_i&=P_i\otimes \mathbb{1}+\mathbb{1}\otimes P_i\, ,
\end{align}
where 
\begin{align}
    \Pi_0&=\ell P_0+\sqrt{1-\ell^2 P^{\mu}P_{\mu}}\, ,\\
    \Pi_0^{-1}&=\frac{\sqrt{1-\ell^2 P^{\mu}P_{\mu}}-\ell P_0}{1-\ell^2{\bf P}^2}\, ,\\
    P^{\mu}P_{\mu}&=-P_0^2+{\bf P}^2=-P_0^2+\sum_iP_iP_i\, .
\end{align}

This way of expressing the generators of translations and its action on two-particle states is called the {\it classical basis} of the $\kappa$-Poincaré algebra \cite{Borowiec:2009vb}. It is named this way because the symmetry generators and, consequently, the mass Casimir operator of the algebra is undeformed for single particle systems, and the modifications are concentrated on the multi-particle sector. An interesting aspect of this  approach is that it admits a nonrelativistic limit that retains traces of the quantum gravity scale, i.e., after recovering the speed of light $c$ withing the symmetry generators and quantum gravity parameter $\ell$, one can consider the non-relativistic (Galilean) limit by taking $c\rightarrow \infty$. In this classical basis, the nonrelativistic limit of the antipode and coproduct takes the form:
\begin{align}
S(P_0) &= -P_0 - i\ell \mathbf{P}^2, \\
    S(P_i) &= -P_i + \ell P_i P_0\, ,\\
    \Delta P_0 &= P_0 \otimes \mathbb{1} + \mathbb{1} \otimes P_0 - i\ell \sum_i P_i \otimes P_i, \\
    \Delta P_i &= P_i \otimes \mathbb{1} + \mathbb{1} \otimes P_i\, .
\end{align}

The form of the coproduct operators leads to a modified, non-Abelian conservation law for energy and momentum. When applied to a composite system $|g\rangle \otimes |h\rangle$, the momentum eigenvalues of the composite system are then given by:
\begin{align}
    g_0 \oplus h_0 &= g_0 + h_0 + \ell \vec{g} \cdot \vec{h}, \label{eq:def_comp} \\
    g_i \oplus h_i &= g_i + h_i,
\end{align}
where $(g_0, g_i)$ represents the energy and momentum (in Cartesian coordinates) of the single-particle state $|g\rangle$ (and similarly for $h$).


\subsection{Lindblad Equation}

The presence of deformed symmetries induces a deformation of the commutator, given by the adjoint action \cite{Arzano:2014cya,Arzano:2022nlo}. This action selects a representative of the vector space ${\cal R}_G$, spanned by the generators of the symmetry algebra $G \in {\cal R}_G$, and acts on the quantum system ${\cal A}_{\cal H}$ defined on a Hilbert space ${\cal H}$ as:
\begin{equation}
    \text{ad}_{\clubsuit}\spadesuit \, :\, {\cal R}_G \times {\cal A}_{\cal H} \rightarrow {\cal A}_{\cal H}, \quad \text{ad}_G O = (\text{id} \otimes S)\Delta G \diamond O,
\end{equation}
where $\text{id}$ is the identity operator, $\Delta G$ is the coproduct operator, $S$ is the antipode map, and $\diamond$ is the operator defined as:
\begin{equation}
    (a \otimes b) \diamond O \doteq a O b.
\end{equation}

When the coproduct operator is of the form $\Delta G = \sum_i a_i \otimes b_i$, then $(\text{id} \otimes S)\Delta G = \sum_i a_i \otimes S(b_i)$, which implies that $\text{ad}_G O = \sum_i a_i O S(b_i)$. If $O = \rho$ is the density matrix of a quantum system and $\Delta G = P_0 \otimes \mathbb{1} + \mathbb{1} \otimes P_0$, then $\text{ad}_{P_0} \rho = [P_0, \rho]$. This means that the evolution equation of a system with Galilean invariance can be written as $i\partial_t \rho = \text{ad}_{P_0} \rho$. For a quantum spacetime with deformed symmetries, the generalization of this evolution equation in a way that preserves the hermiticity of $\rho$ is:
\begin{equation}
    i\partial_t \rho = \frac{1}{2} \left\{ \text{ad}_{P_0} \rho - \left[ \text{ad}_{P_0} \rho \right]^{\dagger} \right\}.
\end{equation}

A straightforward calculation based on the above definition leads to a Lindblad equation for a quantum system with deformed symmetries:
\begin{equation}\label{eq:main_lindblad}
    \partial_t \rho = -i[P_0, \rho] - \frac{\ell}{2} \left( \mathbf{P}^2 \rho + \rho \mathbf{P}^2 - 2 \sum_i P_i \rho P_i \right). 
\end{equation}

This implies that the non-commutative spacetime acts effectively as an environment of an open quantum system. Considering a free nonrelativistic particle, we can describe the elements of the density matrix in the momentum representation as $\rho_{pq}=\langle {\bf p}|\rho|{\bf q}\rangle$, ${\bf p}$ represents the components of the particle's momentum. From the dispersion relation $E(p)=p^2/2m$, the evolution of these elements is
\begin{equation}
    \partial_t \rho_{pq}=\left[-\frac{i}{\hbar}(E(p)-E(q))-\frac{\ell}{2}({\bf p}-{\bf q})^2\right]\rho_{pq}.
\end{equation}
whose solution is
\begin{equation}
    \rho_{pq}(t)=\rho(0)\exp\left[-it\left(E(p)-E(q)\right)-\frac{\ell t}{2}({\bf p}-{\bf q})^2\right]\, .
\end{equation}

We observe that the off-diagonal elements  (${\bf p}\neq {\bf q}$) of the density matrix exponentially decays. The time scale of such decay is called decoherence time, and in this model is given by
\begin{equation}
    \tau_D=\frac{2}{\ell({\bf p}-{\bf q})^2}.
\end{equation}

Since the length scale of quantum gravity should be very small, the decoherence time should be very large and the decoherence is suppressed by quantum gravity deformations. The presence of a Lindblad equation implies that a quantum spacetime acts as a source of fundamental decoherence in quantum systems \cite{Arzano:2022nlo}. In the next section, we shall deal with corrections coming from the Hamiltonian side of the evolution equation, when we are able to separate the quantum system in two sectors, defined by two particles, where the conservation law implies a modified Hamiltonian of the system.


\section{Two-Particle Model}\label{sec:two_part}

The modification of the energy conservation law, given by Eq.~\eqref{eq:def_comp}, leads to the definition of the Hamiltonian for a system of two free particles as:
\begin{equation}\label{eq:Hamiltonian}
    H = H_0 \otimes \mathbb{1} + \mathbb{1} \otimes H_0 + \ell \sum_i P_i \otimes P_i,
\end{equation}
where $H_0$ is the Hamiltonian of a free particle. Notably, the new term in \eqref{eq:Hamiltonian} resembles a momentum-dependent potential that may induce correlations between the particles.

In the following, we will analyze the case of two interacting particles, each occupying a defined energy state.


\subsection{Two-Energy States}

To estimate these effects, we consider a simple case of two particles, each modeled as a two-level system. Individually, each particle can occupy energy states with energies $E_{(g)} = \dfrac{\epsilon - \omega}{2}$ or $E_{(e)} = \dfrac{\epsilon + \omega}{2}$, where $\epsilon > \omega$ ensures positive energies. If the energy composition were undeformed, the Hamiltonian of this system would be:
\begin{equation}
    H_{\text{und}} = \epsilon \mathbb{1} \otimes \mathbb{1} + \frac{\omega}{2} (\sigma_z \otimes \mathbb{1} + \mathbb{1} \otimes \sigma_z),
\end{equation}
where $\sigma_z$ is the $z$-Pauli matrix. The two-particle state would then be described by two indices $|ij\rangle$, where $(ij) = \{g, e\}$ labels whether the first or second particle is in the ground or excited state.

Now, suppose the particles are in a superposition of momenta. Typically, this would not affect the system's description, but since the inner product of momenta influences the energy of the composite system, the direction of each momentum becomes important. For simplicity, we consider a one-dimensional case, which suffices for our purposes. Here, each particle can move in either the positive ($|+\rangle$) or negative direction ($|-\rangle$), meaning the composite system is described by a state $|ij\alpha\beta\rangle$, where $(\alpha, \beta) = \{+, -\}$ labels the direction of the first or second particle.

In our model, each particle can have either positive or negative momentum. This means that the eigenvectors of direction for each particle can be written as a linear combination of energy eigenstates:
\begin{equation}\label{eq:plus_minus}
    |\pm\rangle = \frac{1}{\sqrt{2}} (|g\rangle \pm |e\rangle).
\end{equation}
This approximation is reasonable, as it can be verified by combining energy eigenfunctions of a 1D harmonic oscillator in the momentum representation, showing that this combination is concentrated in the positive or negative branch of momentum.

From the undeformed dispersion relation (since we are at the classical basis), we construct an operator that gives the modulus of the momentum of a two-level state, sharing the same energy eigenstates but with different energies for each level:
\begin{equation}
    |P| = \text{diag}\left( \sqrt{m(\epsilon + \omega)}, \sqrt{m(\epsilon - \omega)} \right).
\end{equation}
To account for the momentum direction, we note that the $x$-Pauli matrix has eigenstates given by Eq.~\eqref{eq:plus_minus}. Thus, the Hamiltonian describing the evolution of this system is:
\begin{align}
    {\cal H} &= \epsilon (\mathbb{1})^4 + \dfrac{\omega}{2} \left( \sigma_{z} \otimes \mathbb{1} \otimes \mathbb{1} \otimes \mathbb{1} + \mathbb{1} \otimes \sigma_{z} \otimes \mathbb{1} \otimes \mathbb{1} \right) + \ell |P| \otimes |P| \otimes \sigma_{x} \otimes \sigma_{x},
\end{align}
where $(\mathbb{1})^4 = \mathbb{1} \otimes \mathbb{1} \otimes \mathbb{1} \otimes \mathbb{1}$, and $\mathbb{1}$ is the $2 \times 2$ identity matrix. This Hamiltonian is a $16 \times 16$ matrix, such that a direct computation shows that it is constituted by four $4 \times 4$ block matrices as follows:
\begin{equation}\label{eq:block_ham}
    {\cal H} =
    \begin{pmatrix}
        H_{11} & 0 & 0 & 0 \\
        0 & H_{-11} & 0 & 0 \\
        0 & 0 & H_{1-1} & 0 \\
        0 & 0 & 0 & H_{-1-1}
    \end{pmatrix},
\end{equation}
where
\begin{widetext}
\begin{equation}\label{eq:hamil}
    H_{ab} =
    \begin{pmatrix}
        \epsilon + \frac{a + b}{2} \omega & 0 & 0 & \ell m \sqrt{(\epsilon + a \omega)(\epsilon + b \omega)} \\
        0 & \epsilon + \frac{a + b}{2} \omega & \ell m \sqrt{(\epsilon + a \omega)(\epsilon + b \omega)} & 0 \\
        0 & \ell m \sqrt{(\epsilon + a \omega)(\epsilon + b \omega)} & \epsilon + \frac{a + b}{2} \omega & 0 \\
        \ell m \sqrt{(\epsilon + a \omega)(\epsilon + b \omega)} & 0 & 0 & \epsilon + \frac{a + b}{2} \omega
    \end{pmatrix},
\end{equation}
\end{widetext}
with $(a, b) = \{-1, 1\}$. Since the evolution of this system involves Lindblad operators, we need to describe the momentum of the composite system. Given that the Lindblad operators are multiplied by $\ell$ in \eqref{eq:main_lindblad} and are expected to contribute only slightly, the momentum composition remains undeformed. The operator representing the sum of the momenta of the two particles is:
\begin{equation}
    {\cal P} = |P| \otimes \mathbb{1} \otimes \sigma_x \otimes \mathbb{1} + \mathbb{1} \otimes |P| \otimes \mathbb{1} \otimes \sigma_x,
\end{equation}
where we account for the possibility that the particles may be moving in different directions. This operator can also be expressed in block form, similar to the Hamiltonian in \eqref{eq:block_ham} as
\begin{equation}\label{eq:block_ham}
    {\cal P} =
    \begin{pmatrix}
        P_{11} & 0 & 0 & 0 \\
        0 & P_{-11} & 0 & 0 \\
        0 & 0 & P_{1-1} & 0 \\
        0 & 0 & 0 & P_{-1-1}
    \end{pmatrix}\, ,
\end{equation}
where
\begin{widetext}
\begin{equation}\label{eq:pab}
    P_{ab} = \sqrt{m}
    \begin{pmatrix}
        0 & \sqrt{\epsilon + a \omega} & \sqrt{\epsilon + b \omega} & 0 \\
        \sqrt{\epsilon + a \omega} & 0 & 0 & \sqrt{\epsilon + b \omega} \\
        \sqrt{\epsilon + b \omega} & 0 & 0 & \sqrt{\epsilon + a \omega} \\
        0 & \sqrt{\epsilon + b \omega} & \sqrt{\epsilon + a \omega} & 0
    \end{pmatrix}.
\end{equation}
\end{widetext}

Although $P_{ab}$ has not an $X$-shape, the matrix $(P_{ab})^2$ has this form, just like the Hamiltonian \eqref{eq:hamil}. As this operator appears explicitly in the Lindblad equation \eqref{eq:main_lindblad}, this suggests that a natural ansatz for the density matrix consists also in a block-diagonal matrix $\rho = \text{diag}\left( \rho_{11}, \rho_{-11}, \rho_{1-1}, \rho_{-1-1} \right)$, where each $\rho_{ab}$ is a $4\times 4$ $X$-shaped matrix. In fact, assuming this ansatz, the last term of the Lindblad equation \eqref{eq:main_lindblad} $-2{\cal P} \rho {\cal P}$ also assumes a block diagonal form of $X$-shaped matrices labeled by parameters $(a,b)$ running $\{-1,1\}$.

As a consequence of these observations, we can reduce the problem of handling the system of differential equations described by a $16\times 16$ matrix to four sets of systems of differential equations given by $4\times 4$ matrices, labeled by the parameters $(a,b)$. Specifically, for each block of matrices, obeys a Lindblad equation of the form:
\begin{equation}\label{eq:lindblad}
    \partial_t \rho_{ab} + i \left[  H_{ab}, \rho_{ab} \right] + \frac{\ell}{2} \left( P_{ab}^2 \rho_{ab} + \rho_{ab} P_{ab}^2 - 2 P_{ab} \rho_{ab} P_{ab} \right) = 0.
\end{equation}

Since the particles can move in any direction, there is a $25\%$ probability for the system to be in one of the four possible configurations of parameters $(a, b) = \{-1, 1\}$. Therefore, the density matrix used to compute correlations must be averaged as:
\begin{equation}\label{eq:aver_density}
    \bar{\rho} = 0.25 \left(\rho_{-1-1}+\rho_{-11}+\rho_{1-1}+\rho_{11}\right).
\end{equation}
with elements

\begin{equation}
   \bar{\rho}_{ij}\doteq [\bar{\rho}]_{ij} = 0.25 \left([\rho_{-1-1}]_{ij}+[\rho_{-11}]_{ij}+[\rho_{1-1}]_{ij}+[\rho_{11}]_{ij}\right)
\end{equation}
To analyze this state, we consider solutions to the Lindblad equation, following an ansatz similar to the one adopted in \cite{Carrion:2024lap}, assuming the initial condition:
\begin{equation}\label{eq:init_cond}
    |\Psi\rangle = \sin \theta |eg\rangle + \cos \theta |ge\rangle,
\end{equation}
which represents an unentangled state when $\theta = 0$ and a maximally entangled state (Bell state) when $\theta = \pi/4$. Thus, the parameter $\theta$ allows us to control the initial level of entanglement. The corresponding averaged matrix for this initial conditions is
\begin{eqnarray}\label{eq:init_cond2}
  \bar{\rho}(t=0)=  \left(
\begin{array}{cccc}
 0 & 0 & 0 & 0 \\
 0 & \cos ^2(\theta ) & \sin (\theta ) \cos (\theta ) & 0 \\
 0 & \sin (\theta ) \cos (\theta ) & \sin ^2(\theta ) & 0 \\
 0 & 0 & 0 & 0 \\
\end{array}
\right).
\end{eqnarray}

In appendix \ref{sec:app1}, we solve this Lindblad equation analytically, and derive its asymptotic behavior for late times.

\section{Quantum Correlations}\label{sec:quant_corr}

Quantum spacetime is typically responsible for introducing dissipation in the form of decoherence in quantum systems (see, for instance, \cite{Petruzziello:2020wkd}). However, a unique signature of the deformation of the relativity principle is the presence of modified composition laws for energy and momentum. For single-particle states, decoherence emerges as a consequence of the modified evolution equation \eqref{eq:main_lindblad}, as discussed in \cite{Arzano:2022nlo}. However, for multiple particles, the introduction of interaction-like terms that depend on the momenta of each particle may introduce correlations that initially compete with decoherence effects. In this section, we will consider four quantifiers of correlations: concurrence, as a measure of entanglement; the $l_1$-norm of coherence, quantum mutual information, and local quantum Fisher information. 


\subsection{Concurrence}

Entanglement is the most profound quantum correlation between systems and was first highlighted in the work by Einstein, Podolsky, and Rosen \cite{PhysRev.47.777}, which aimed to point out an incompleteness in the quantum description of the world. The modern definition of entanglement dates back to the works by Werner \cite{PhysRevA.40.4277}, where this concept was extended to mixed states. Today, entanglement plays a fundamental role in quantum cryptography \cite{PhysRevLett.67.661}, quantum computing \cite{PhysRevLett.69.2881}, and quantum teleportation \cite{PhysRevLett.70.1895}.

A useful quantifier of entanglement for a two-qubit state, described by an $X$-shaped matrix, is the concurrence. For a $4 \times 4$ matrix, it reads \cite{Wootters:1997id,Hill:1997pfa}:
\begin{equation}
    \mathcal{C} = 2 \max \left\{ |\bar{\rho}_{23}| - \sqrt{\bar{\rho}_{11} \bar{\rho}_{44}}, |\bar{\rho}_{14}| - \sqrt{\bar{\rho}_{22} \bar{\rho}_{33}}|, 0 \right\}, \label{eq:conc}
\end{equation}
where the density matrix used is the averaged one described in \eqref{eq:aver_density}. The system of differential equations \eqref{eq:lindblad} for the initial condition \eqref{eq:init_cond} was solved analytically for arbitrary $\theta$ in section \ref{sec:app1}. The asymptotic behavior of the system is quoted below:
\begin{align}
    &\lim_{t\rightarrow \infty}\rho_{11}=\lim_{t\rightarrow \infty}\rho_{22}=\lim_{t\rightarrow \infty}\rho_{33}=\lim_{t\rightarrow \infty}\rho_{44}=\frac{1}{4},\\
    &\lim_{t\rightarrow \infty}\rho_{14}=\lim_{t\rightarrow \infty}\rho_{23}=\frac{\sin({\theta})\cos(\theta)}{2}=\frac{\sin{(2\theta)}}{4}.\label{eq:assymp}
\end{align}

We verify that the diagonal elements of the density matrix eventually larger than the off-diagonal ones when $\theta\neq \pi/4$. This means that in a finite time, the concurrence becomes null. For the case where $\theta=\pi/4$, this happens asymptotically.

\begin{figure}[h]
    \centering
    \includegraphics[scale=0.6]{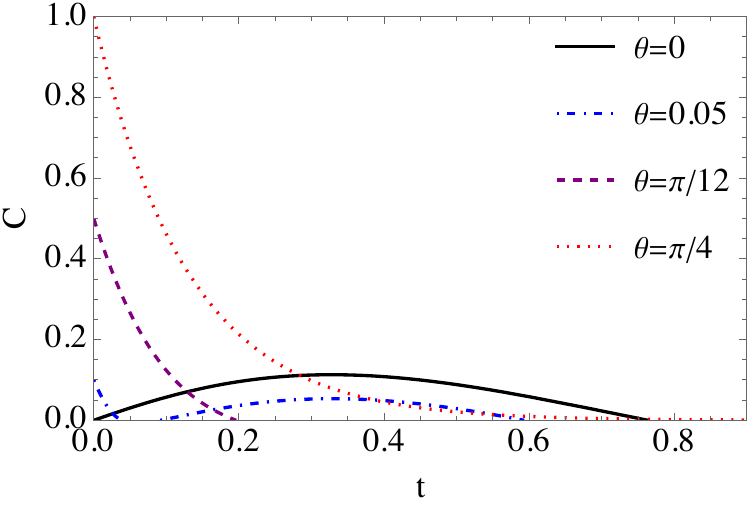}
\caption{Concurrence for different initial conditions determined by $\theta$. The parameters used are $\ell = 1$, $m = 1$, $\epsilon = 1$, and $\omega = 0.5$.}
    \label{fig:conc}
\end{figure}

In Fig.~\ref{fig:conc}, we observe the concurrence for different initial conditions. The qualitative behavior of the quantities discussed in this paper does not change significantly for different values of the parameter $\ell$. Only the scale in which they occur is modified. Therefore, without loss of generality, we consider $\ell$ with the same order of magnitude of the other parameters of the model. We see that a system of two particles can be prepared in an unentangled state ($\theta = 0$) and become entangled over time due to the modified energy conservation in deformed relativity (black, solid curve). By increasing the angle to $\theta = 0.05$, we see that although the state is initially partially entangled, decoherence effects reduce the concurrence to zero. However, the modified composition law effects are still strong enough to reentangle the system before it becomes completely unentangled (blue, dash-dotted curve). For $\theta = \pi/12$, the system starts more entangled before experiencing a drop in this correlation (purple, dashed curve). This behavior repeats for the initial Bell state ($\theta = \pi/4$) (red, dotted curve). We considered unrealistically large values of the parameters for the sake of clarity of the figures, but this does not affect the main results. In fact, we are considering a value of the $\ell$-parameter of the same order of the other parameters in this plot, but in reality, it should be much smaller, since it is related to the Planck length scale of quantum gravity.

Notice that such mechanism is conceptually different than the one from the gravitationally-induced entanglement procedure \cite{Bose:2017nin,Marletto:2017kzi}. In their case, the Newtonian potential and a superposition of position states (which creates an effective superposition of gravitational interactions) is responsible for creating entanglement between the two massive particles. In our case, we interpret that after an interaction, considering that the particles are in a momentum superposition, the initially separate state of the two quantum particles, becomes entangled due to the non-commutative properties of spacetime. In the former case, it has been argued that the Newtonian interaction creates a communication channel that violates the Local Operation and Classical Communication (LOCC) theorem, which, as a consequence, therefore implies that the gravitational interaction is of quantum nature in essence. In our case, we consider that the particles are free, i.e., there is no interaction potential that serves as a channel for communication between them. The term $\vec{p}_1\cdot\vec{p}_2$ that corrects the undeformed Hamiltonian is not, in fact, an interaction term due to some force. Although it, in a broad sense, ``interacts'' the particles through a scalar product, its presence here is purely kinematical in nature.

The extra, non-local term that deforms the energy conservation is introduced by the same mechanism that creates decoherence between the particles, i.e., the underlying quantum algebra that is responsible for the preservation of the relativity principle at the Planck scale. In fact, as we verified in Fig.\ref{fig:conc}, the time scale of the decoherence is the same as the entanglement generation. This entanglement generation is due to the apparent non-locality of this extra term in the conservation law. Non-local terms can, in fact, create entanglement and circumvents the LOCC theorem \cite{Fragkos:2022tbm}. 

The non-local nature of modified conservation laws in corrections of classical systems due to quantum gravity has been a matter of debate in the past years \cite{Hossenfelder:2010tm}, which has led to the development of the Principle of Relative Locality \cite{Amelino-Camelia:2011lvm}, that states that the locality of interactions is an observer-dependent statement, that can be effectively described via modified composition laws. While the Relative Locality framework considers energy-momentum conservation using classical mechanics tools, we are exploring the possible non-local behavior from a quantum mechanical perspective (due to entanglement). Although entanglement emerges in this approach, we also have the presence of decoherence effects, that naturally suppresses such behavior to a limited time window.

If we consider the product state determined by $\theta=0$, then, from \cite{Yang:2017llj}, the time scale that governs the entanglement of this system is given by $T_{\text{ent}} = (\ell \Delta P^2)^{-1}$, where $\Delta P$ is the standard quantum uncertainty of the momentum. If we average the uncertainty of the observable as $\Delta P = 0.25 \sum_{ab} \Delta P_{ab}$ and express the result by restoring the speed of light and Planck's constant, we find:
\begin{equation}\label{eq:ent_time}
    T_{\text{ent}} = \frac{\hbar E_{\text{QG}}}{2mc^2\epsilon},
\end{equation}
where $E_{\text{QG}} = \ell^{-1}$ is the quantum gravity energy scale that measures deviations from standard symmetries. Notice that this expression has an energy dependence similar to the decoherence time found in \cite{Arzano:2022nlo}, $\tau_D \sim 2E_{\text{QG}}/p^2 \sim E_{\text{QG}}/(m E)$. For the parameters used in Fig.~\ref{fig:conc}, $T_{\text{ent}}$ is of the order of $0.5$ in natural units, which is consistent with Fig.~\ref{fig:conc}. To understand this, let us suppose that we prepare such an unentangled superposition of two free particles. The uncertainty in the momentum of each particle translates into the time that it takes to observe the entanglement generation between them.

If the energy scale of the particles is $\epsilon$, then $T_{\text{ent}} \epsilon \sim \hbar E_{\text{QG}}/(mc^2)$. For instance, consider a pair of ions \cite{Krutyanskiy:2022yoa} with a mass of the order $m = 3.7\times 10^{10}\, \text{eV}$, where the energy scale of electronic transitions is of the order $\epsilon \sim 1\, \text{eV}$, the Planck energy $E_{\text{QG}} = E_{\text{Planck}} = 1.2\times 10^{28}\, \text{eV}$ and $\hbar=6.6\times 10^{-16}\text{eV}\cdot\text{s}$. We can estimate the entanglement time scale $T_{\text{ent}} =214\, \text{s}\approx 3.6\, \text{min}$. If an entangled pair of ions has approximately $100$ elementary particles and supposing that the quantum gravity energy scale grows linearly with the number of particles \cite{Hossenfelder:2014ifa,Amelino-Camelia:2013fxa}, the entanglement time becomes of the order of hours. This means that it would be necessary to preserve the quantum properties of this superposition of two particle system along this time scale in order to probe entanglement generation due to the uncertainty in their momenta. Besides that, if the quantum gravity scale $E_{\text{QG}}$ grows with the number of elementary particles, then the entanglement time scale grows too. Notice that this time scale is larger than the one typically considered in the gravitationally induced entanglement proposals \cite{Bose:2017nin,Marletto:2017kzi} of $\sim 1$ s (also for different experimental setups, of course). But if one is capable of designing such an experiment, it would be possible at least to set interesting constraints on the quantum gravity scale not so far from the Planck scale.

An interesting opportunity could be achieved when we consider long lived qubits, like trapped ions. For instance, in \cite{wang2020single} it is reported coherence time exceeding one hour for a single qubit. Coherence time of system of qubits can serve as laboratory tests for deformations of the conservation of energy and deformed Hamiltonians. Of course, in models in which $\Delta P$ can be enlarged would reduce the entangling time, which could be interesting for experimental tests.

Curiously, if we make $\epsilon=mc^2$ in \eqref{eq:ent_time}, and we compare it with the entanglement time scale of the gravitationally induced setting $T_{\text{GIE}}=\hbar L/(Gm^2)$ (where $L$ is the scale of the relative distance between the particles of the system), a straightforward calculation shows that $L=\ell$. This means that at the Planckian regime, the gravitational interaction between the particles involved in the scenario discussed in this paper cannot be ignored.


\subsection{The $l_1$-Norm of Coherence}

Coherence is an important ingredient in various areas of physics, such as quantum optics, quantum information, solid-state physics, and biology \cite{li2012witnessing}. The first axiomatic approach to quantifying coherence was proposed in \cite{Aberg:2006xcv}, and since then, several approaches have been developed. For a review, we refer the reader to \cite{Streltsov:2016iow}. In this paper, we use the $l_1$-norm of coherence, introduced by \cite{Baumgratz:2013ecx}, which is a quantifier based on matrix norms and can be straightforwardly applied to the type of density matrix we are analyzing.

For an $X$-shaped density matrix, it reads:
\begin{equation}\label{eq:l1norm}
    C_{l_1} = \sum_{i \neq j} |\bar{\rho}_{ij}| = 2(|\bar{\rho}_{14}| + |\bar{\rho}_{23}|),
\end{equation}
which simply measures the magnitude of the off-diagonal elements of $\bar{\rho}$. 

\begin{figure}
    \centering
    \includegraphics[scale=0.6]{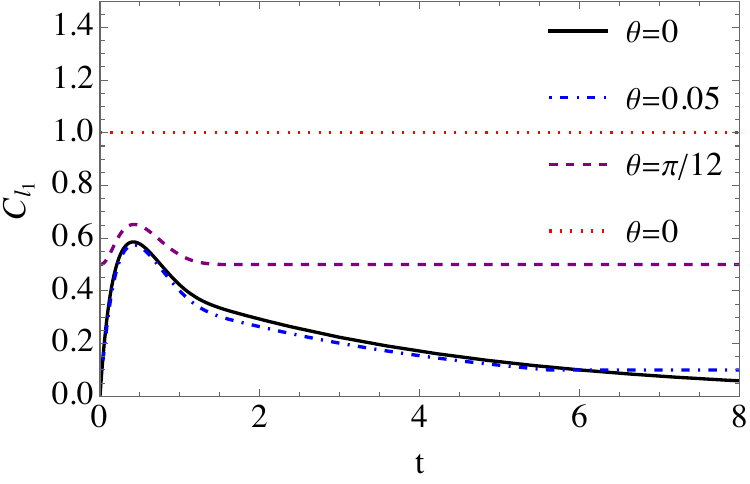}
    \caption{The $l_1$-norm of coherence for different initial conditions determined by $\theta$. The parameters used are $\ell = 1$, $m = 1$, $\epsilon = 1$, and $\omega = 0.5$.}
    \label{fig:coh}
\end{figure}

In Fig.~\ref{fig:coh}, we see that for $\theta < \pi/4$, the Hamiltonian creates correlations and initially dominates over decoherence effects, sharing the behavior of the concurrence discussed in the previous subsection. Over time, the Lindbladian terms eventually balance the deformed composition law contribution. Asymptotically, we observe a perfect balance between the creation of correlations due to the Hamiltonian and decoherence effects due to the Lindblad equation, which preserves the coherence of the system when $\theta\neq0$. This is due to the asymptotic behavior of the density matrix elements shown in appendix \ref{sec:app1}. Notably, for an initial Bell state ($\theta = \pi/4$), since the system begins maximally coherent, these two contributions balance each other from the outset. In fact, the open quantum spacetime also creates correlations in this system, as we will discuss in Section \ref{sec:diff_l}, meaning that while it dissipates entanglement, it helps maintain other correlations between the particles for initially partially entangled states.

\begin{figure}[htbp]
    \centering
    \begin{subfigure}[b]{0.45\textwidth}
        \centering
        \includegraphics[width=\textwidth]{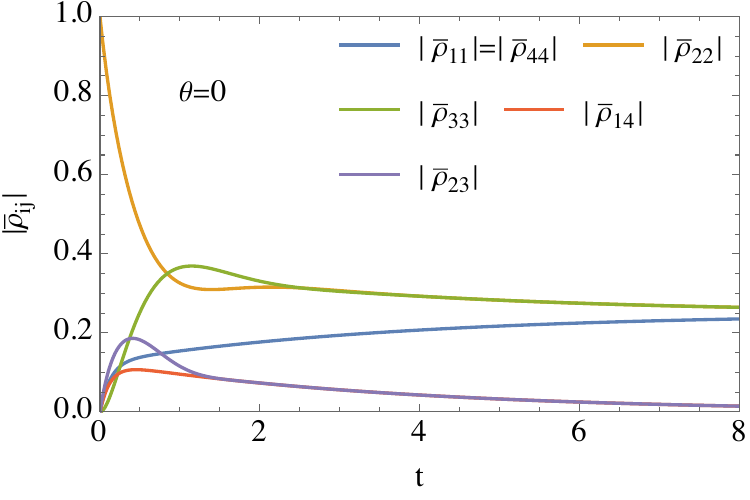}
        \caption{Behavior of the average density matrix for $\theta=0$.}
        \label{fig:dens1}
    \end{subfigure}
    \hfill
    \begin{subfigure}[b]{0.47\textwidth}
        \centering
        \includegraphics[width=\textwidth]{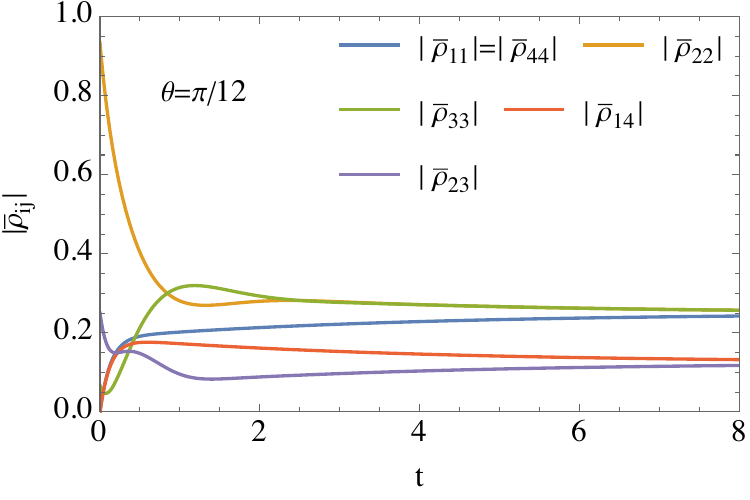}
        \caption{Behavior of the average density matrix for $\theta=\pi/12$.}
        \label{fig:dens2}
    \end{subfigure}
    \caption{Behavior of the average density matrix elements $\bar{\rho}_{ij}$ for the initial condition $\theta=0$ and $\theta=\pi/12$. We assume parameters $\ell=1$, $m=1$, $\epsilon=1$, $\omega=0.5$.}
    \label{fig:main}
\end{figure}

We observe that such behavior is due to the survival of the off-diagonal elements of the density matrix as can be seen in Eq.\eqref{eq:assymp}. For an initially unentangled state, the off-diagonal elements vanish in time as can be seen in Fig. \ref{fig:dens1}. In this case, we see that the off-diagonal elements $|\bar{\rho}_{14}|$ (orange curve) and $|\bar{\rho}_{23}|$ (purple curve) asymptotically survive and assume the same value. Also notice in Fig. \ref{fig:dens1} that, at initial times, the element $|\bar{\rho}_{23}|$ (purple curve) is indeed larger than $\bar{\rho}_{11}\bar{\rho}_{44}$ (blue curve), which is the reason why there is an initial entanglement due to the modified composition law. Also notice in Fig. \ref{fig:dens2} that for a initially partially entangled state, the off-diagonal elements grow again and approach each other according to \eqref{eq:assymp}.

This behavior is not entirely new in condensed matter systems. For instance, a similar phenomenon can be observed in \cite{Carrion:2024lap}, which considers a spin-$1/2$ Ising-$XYZ$ chain model. There, one also observes the disappearance and resurgence of entanglement and coherence, with the stabilization of quantum correlations over time.


\subsection{Quantum Mutual Information}\label{sec:disc}

An important quantifier of quantum correlations beyond entanglement is the Quantum Discord. By calculating it, we can determine whether the balance between the Hamiltonian and the Lindbladian dissipates all quantum correlations or only entanglement. The definition of quantum discord uses the concept of {\it mutual information} between two random variables \cite{Streltsov2015,PhysRevLett.88.017901}, which is the correlations that we will consider in this subsection. For a system consisting of two subsystems, say $A$ and $B$ (e.g., the two particles considered in this paper), the density matrix is written as $\rho^{AB}$, which in our case is $\bar{\rho}$. The quantum mutual information is given by:
\begin{equation}\label{eq:Iq}
    I(\rho^{AB}) = S(\rho^A) + S(\rho^B) - S(\rho^{AB}),
\end{equation}
where $S(\rho) = -\text{Tr}[\rho \log_2 \rho]$ is the von Neumann entropy, which measures the uncertainty of a quantum state $\rho$.

The quantity $\rho^A$ is the reduced density operator of subsystem $A$, defined as:
\begin{equation}
    \rho^A = \text{Tr}_B[\rho^{AB}],
\end{equation}
where $\text{Tr}_B$ is the partial trace over subsystem $B$, given by:
\begin{equation}
    \text{Tr}_B[\rho^{AB}] = \sum_{k=1}^d ({\mathbb 1}_A \otimes \langle k_B|) \rho^{AB} ({\mathbb 1}_A \otimes |k\rangle_B),
\end{equation}
where $d$ is the dimension of subsystem $B$, ${\mathbb 1}_A$ is the identity operator of subsystem $A$ (in our case, $\text{diag}(1,1)$), and $\{|k\rangle_B\}$ is an orthonormal basis of subsystem $B$ \cite{maziero2017computing,Watrous:2018zil}. In our case, $d = 2$, and we consider the computational basis $\{|k\rangle_B\} = \{|0\rangle, |1\rangle\}$. For a $4 \times 4$ matrix with elements $[\rho^{AB}]_{ij} = \rho_{ij}$, we have:
\begin{equation}
    \rho^A = \text{Tr}_B(\rho^{AB}) = \begin{pmatrix}
        \rho_{11} + \rho_{33} & \rho_{12} + \rho_{34} \\
        \rho_{21} + \rho_{43} & \rho_{22} + \rho_{44}
    \end{pmatrix},
\end{equation}
\begin{equation}
    \rho^B = \text{Tr}_A(\rho^{AB}) = \begin{pmatrix}
        \rho_{11} + \rho_{22} & \rho_{13} + \rho_{24} \\
        \rho_{31} + \rho_{42} & \rho_{33} + \rho_{44}
    \end{pmatrix}.
\end{equation}

This procedure provides a way to derive the state of $A$ by tracing out the degrees of freedom of $B$, and vice versa. It ensures the correct measurement statistics for measurements made on subsystems $A$ and $B$ \cite{Nielsen:2012yss}.

For a classical system, there is an equivalent way to define mutual information. The Shannon entropy is the classical counterpart of the von Neumann entropy. Consider a classical system described by two random variables $X$ and $Y$, where $p_x$ is the probability that $X$ takes the value $x$, $p_y$ is the probability that $Y$ takes the value $y$, and $p_{x,y}$ is the probability that a measurement of $(X, Y)$ yields $(x, y)$. The Shannon entropy for $X$ is given by $H(X) = -\sum_x p_x \log_2 p_x$.

From Bayes' rule, the probability that $X$ takes the value $x$, given that $Y$ takes the value $y$, is $p_{x|y} = p_{xy} / p_y$. This allows us to write:
\begin{align}
    H(X, Y) = -\left(\sum_y p_y \log_2 p_y\right) \left(\sum_x p_{x|y}\right)- \sum_y p_y \left(\sum_x p_{x|y} \log_2 p_{x|y}\right) = H(Y) + H(X|Y),
\end{align}
where $H(X|Y) = \sum_y H(X|y)$ is the conditional entropy, representing the amount of information needed to describe the outcome of $X$ given the outcome of $Y$. Thus, there are two equivalent ways, $I$ and $J$, to describe mutual information for a classical system:
\begin{align}
    I(X, Y) &= H(X) + H(Y) - H(X, Y), \label{eq:I} \\
    J(X, Y) &= H(X) - H(X|Y). \label{eq:J}
\end{align}

A quantum analog of the conditional entropy for a bipartite system $\rho^{AB}$ was introduced by Ollivier and Zurek \cite{PhysRevLett.88.017901}:
\begin{equation}
    S(A|\{\Pi_i^B\}) = \sum_i p_i S(\rho_i^A),
\end{equation}
where $\{\Pi_i^B\}$ are orthogonal measurement operators on subsystem $B$, and $p_i = \text{Tr}[\Pi_i^B \rho^{AB}]$ is the probability of obtaining outcome $i$. Additionally, $\rho_i^A = \text{Tr}_B[\Pi_i^B \rho^{AB}] / p_i$ is the post-measurement state of subsystem $A$. Thus, the quantum version of $J$ defined in \eqref{eq:J} is:
\begin{equation}\label{eq:q_J}
    J(\rho^{AB})_{\{\Pi_i^B\}} = S(\rho^A) - S(A|\{\Pi_i^B\}),
\end{equation}
which depends on the measurement operator $\Pi_i^B$ and represents the amount of information gained about $A$ by measuring $B$. The quantum discord is the difference between the expressions \eqref{eq:Iq} and \eqref{eq:q_J}, minimized over all von Neumann measurements \footnote{A general quantum measurement is described by a collection $\{E_i\}$ of measurement operators that satisfy the completeness equation: $\sum_i E^{\dagger}_i E_i = {\mathbb 1}$. For a projective measurement, the operators $E_i$ are orthogonal projectors: $E_i E_j = \delta_{ij} E_i$. A von Neumann measurement is a special type of projective measurement where the measurement operators $E_i$ are orthogonal projectors with rank one (i.e., their image is a column matrix).}. It is not a simple task to minimize this difference and several works have been dedicated to finding analytical expressions for this quantifier \cite{2018RPPh...81b4001B,chen2011quantum,2010PhRvA..81d2105A,2011PhRvA..83e2108G}. To analyze study correlations of the discord type, we will consider the Local Quantum Fisher Information in the next subsection. Below, we will describe the quantum mutual information between the two particles as a quantifier of correlations.  In Appendix \ref{sec:app2}, we express this quantity for any $X$-state.

\begin{figure}
    \centering
    \includegraphics[scale=0.6]{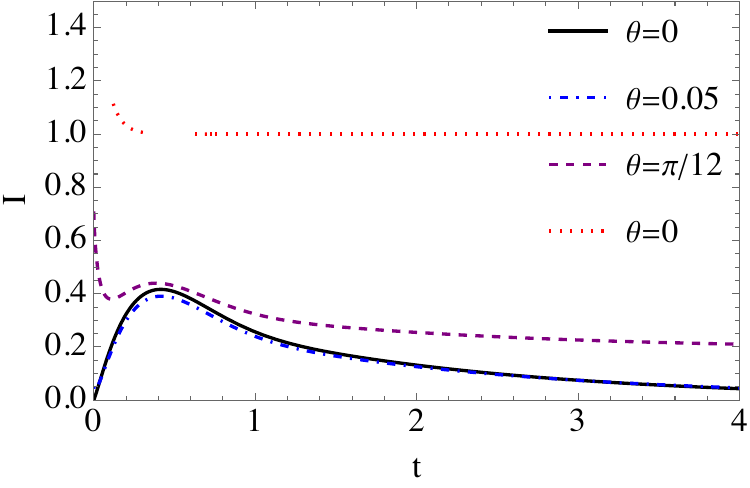}
    \caption{Quantum Mutual Information for different initial conditions determined by $\theta$. We assume parameters $\ell=1$, $m=1$, $\epsilon=1$, $\omega=0.5$.}
    \label{fig:qdisc}
\end{figure}

We depict the mutual information for different initial conditions in Fig.~\ref{fig:qdisc}. It also stabilizes over very long times, indicating that while entanglement degrades over time, other quantum correlations between the two particles persist.

\begin{figure}[h]
    \centering
    \includegraphics[scale=0.6]{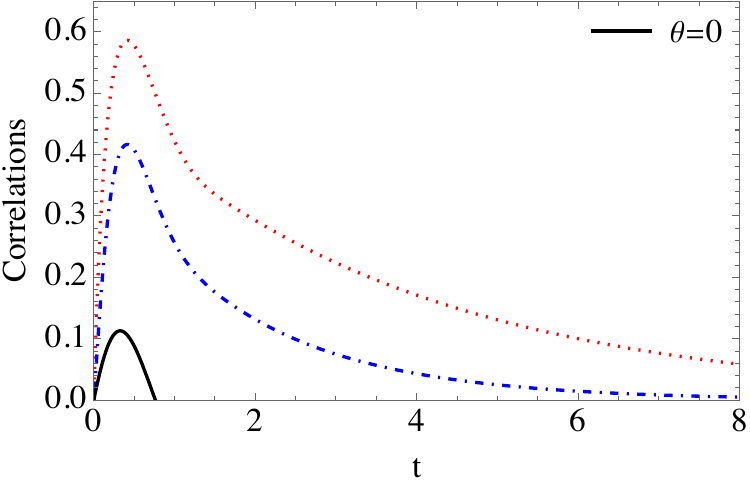}
    \caption{Comparison of the different quantifiers for coherence, quantum mutual information and entanglement studied in the paper for $\theta = 0$. The red/dotted curve represents the $l_1$-norm of coherence. The blue/dash-dotted curve represents quantum mutual information. The black/solid curve represents concurrence.}
    \label{fig:corr-zero}
\end{figure}

We also depict the behavior of the three correlations discussed in this paper so far for an initially unentangled state in Fig.~\ref{fig:corr-zero}. It is evident that there is a hierarchy between coherence, quantum mutual information and entanglement.

\subsection{Local Quantum Fisher Information}

The Quantum Fisher Information (QFI) is a fundamental concept in quantum estimation theory, which measures how much information a quantum state carries about a parameter to be estimated. It is related to the maximum precision with which we can estimate a parameter through the quantum Cramér-Rao bound \cite{slaoui2019comparative,Silva:2025vmu,kim2018characterizing,Dahbi:2022uyi,Lian:2021xjw,Yurischev:2023jrd}.

In many applications of quantum mechanics and quantum information, we aim to estimate a parameter $\phi$ that affects a quantum state $\rho(\phi)$. This parameter can represent, for example: a magnetic field applied to a quantum system, an evolution time of a quantum state or a phase shift in an optical system.
The idea is to measure a quantum system to obtain as much information as possible about $\phi$.

The Quantum Fisher Information is a generalization of classical Fisher Information to quantum states. The formal definition is:
\begin{equation}
    F_Q(\phi) = \text{Tr} \left( \rho(\phi) L_{\phi}^2 \right)\, ,
\end{equation}
where $L_\phi$ is the quantum score operator, also called the Helstrom operator, defined implicitly by the Lyapunov equation:
\begin{equation}
    \frac{d\rho(\phi)}{d\phi} = \frac{1}{2} (\rho(\phi) L_{\phi} + L_{\phi} \rho(\phi)).
\end{equation}

The operator $L_\phi$ represents the best choice of observable for estimating the parameter $\phi$.

The QFI is widely used in various areas of physics and quantum information, such as: quantum metrology, for improving the precision of sensitive measurements, such as interferometry and quantum sensors; quantum communication, for optimizing the transmission and detection of signals encoded in quantum states; quantum computing, for optimizing algorithms based on phase estimation and entangled states, quantum thermodynamics, for understanding the efficiency of quantum heat engines.

In particular, when the parametric states $\rho_\phi$ can be obtained from an initial probe state $\rho$ subjected to a unitary transformation $U_\phi = e^{iH\phi}$ dependent on $\phi$ and generated by a Hermitian operator $H$, that is, $\rho_\phi = U_\phi^\dagger \rho U_\phi$. In this case, the Quantum Fisher Information $F(\rho_\phi)$, which we denote by $F(\rho, H)$, is given by:

\begin{equation}
F(\rho, H) = \sum_{i \neq j} \frac{(p_i - p_j)^2}{p_i + p_j} |\langle \psi_i | H | \psi_j \rangle|^2\, . 
\end{equation}
Here, we use the spectral decomposition of $\rho$, that is,

\begin{equation}
\rho = \sum_i p_i |\psi_i\rangle \langle \psi_i|,
\end{equation} 
with $p_i \geq 0$ and $\sum_i p_i = 1$\, .
The main terms here are:
\begin{itemize}
    \item $p_i$: eigenvalues of the density matrix $\rho$, obtained from its spectral decomposition $\rho = \sum_i p_i |\psi_i\rangle \langle \psi_i|$.
    \item $|\psi_i\rangle$: eigenvectors corresponding to $\rho$.
    \item $H$: Hermitian operator associated with the unitary evolution of the system.
    \item $\frac{(p_i - p_j)^2}{p_i + p_j}$: weight that quantifies the difference between the eigenvalues of the quantum state $\rho$.
    \item $|\langle \psi_i | H | \psi_j \rangle|^2$: matrix element of $H$ between the eigenvectors $|\psi_i\rangle$ and $|\psi_j\rangle$.
\end{itemize}
The important point here is that the QFI is independent of $\phi$. The phase estimation sensitivity, for any type of measurement, is limited by the Crámer-Rao bound $(\Delta\phi)^2\geq 1/F(\rho,H)$ \cite{2014JPhA...47P4006T,1976quantum,holevo2011probabilistic}. Therefore, this quantity provides a lower bound on the precision of the estimation. The knowledge of this limitation is of fundamental importance for quantum metrology, when using quantum mechanics to describe physical systems. For instance, considering the case in which one has an unknown parameter $\alpha$ that controls the strength of a given Hamiltonian that acts on a system during a known time $t$, the unitary evolution of the system is determined by the operator $t\alpha H=\phi H$ for a known observable $H$. Since this argument can be applied for different operations, like rotations through the angular momentum, for example, this estimator is of fundamental importance for several areas of physics that employ different techniques, like interferometry \cite{2014JPhA...47P4006T}.

Now, consider a bipartite quantum state $\rho_{AB}$ of dimension $2 \times d$ in the Hilbert space $H = H_A \otimes H_B$. We assume that the dynamics of the first part is governed by the local phase shift transformation $e^{-i\phi H_A}$, where $H_A \equiv H_a \otimes I_B$ is the local Hamiltonian. In this case, the QFI reduces to the Local Quantum Fisher Information (LQFI), given by:

\begin{equation}
F(\rho, H_A) = \text{Tr}(\rho H_A^2) - \sum_{i \neq j} \frac{2 p_i p_j}{p_i + p_j} |\langle \psi_i | H_A | \psi_j \rangle|^2\, .
\end{equation} 

The Local Quantum Fisher Information (LQFI) was introduced to deal with quantum correlations of the discord type. This quantifier allows a deeper understanding of how quantum correlations influence metrological precision. It has desirable properties that any good quantifier of quantum correlations should satisfy:
\begin{itemize}
    \item It is non-negative and vanishes for bipartite states without discord (classically correlated states).
    \item It is invariant under local unitary operations.
    \item For pure states, it coincides with the geometric discord.
\end{itemize}
The goal of choosing the local Hamiltonian was to capture the quantum correlations between the subsystems. If the system is purely classical, then the LQFI will be zero, as a local observer does not detect quantum effects.

The quantification of quantum correlations in terms of the Local Quantum Fisher Information $Q(\rho)$ is defined as the minimization of the QFI over all local Hamiltonians $H_A$ acting on part $A$:

\begin{equation}
Q(\rho) = \min_{H_A} F(\rho, H_A)\, .
\end{equation} 

The general form of a local Hamiltonian is

\begin{equation}
H_A = \vec{\sigma} \cdot \vec{r}, \nonumber
\end{equation} 
with $|\vec{r}| = 1$ and $\vec{\sigma} = (\sigma_x \equiv \sigma_1, \sigma_y \equiv \sigma_2, \sigma_z \equiv \sigma_3)$ being the usual Pauli matrices.

It can be verified that

\begin{equation}
Tr(\rho H_A^2) = 1, \nonumber
\end{equation} 
and the second term in equation (2) can be rewritten as:

\begin{equation}
\sum_{i \neq j} \frac{2 p_i p_j}{p_i + p_j} |\langle \psi_i | H_A | \psi_j \rangle|^2 = \vec{r}^\dagger M \vec{r}, \nonumber
\end{equation} 
where the elements of the symmetric matrix $M$ of dimension $3 \times 3$ are given by:

\begin{equation}\label{eq:lqf1}
M_{lk} = \sum_{i \neq j} \frac{2 p_i p_j}{p_i + p_j} \langle \psi_i | \sigma_l \otimes I_B | \psi_j \rangle \langle \psi_j | \sigma_k \otimes I_B | \psi_i \rangle.
\end{equation} 

To minimize $F(\rho, H_A)$, it is necessary to maximize the quantity $\vec{r}^\dagger M \vec{r}$ over all unit vectors $\vec{r}$. The maximum value of this expression corresponds to the largest eigenvalue of the matrix $M$.

Therefore, the minimum value of the Local Quantum Fisher Information $Q(\rho)$ is given by:

\begin{equation}
Q(\rho) = 1 - \lambda_{\max}(M), \nonumber
\end{equation} 
where $\lambda_{\max}$ represents the largest eigenvalue of the symmetric matrix $M$ defined in equation \eqref{eq:lqf1}.

\begin{figure}
    \centering
    \includegraphics[scale=0.4]{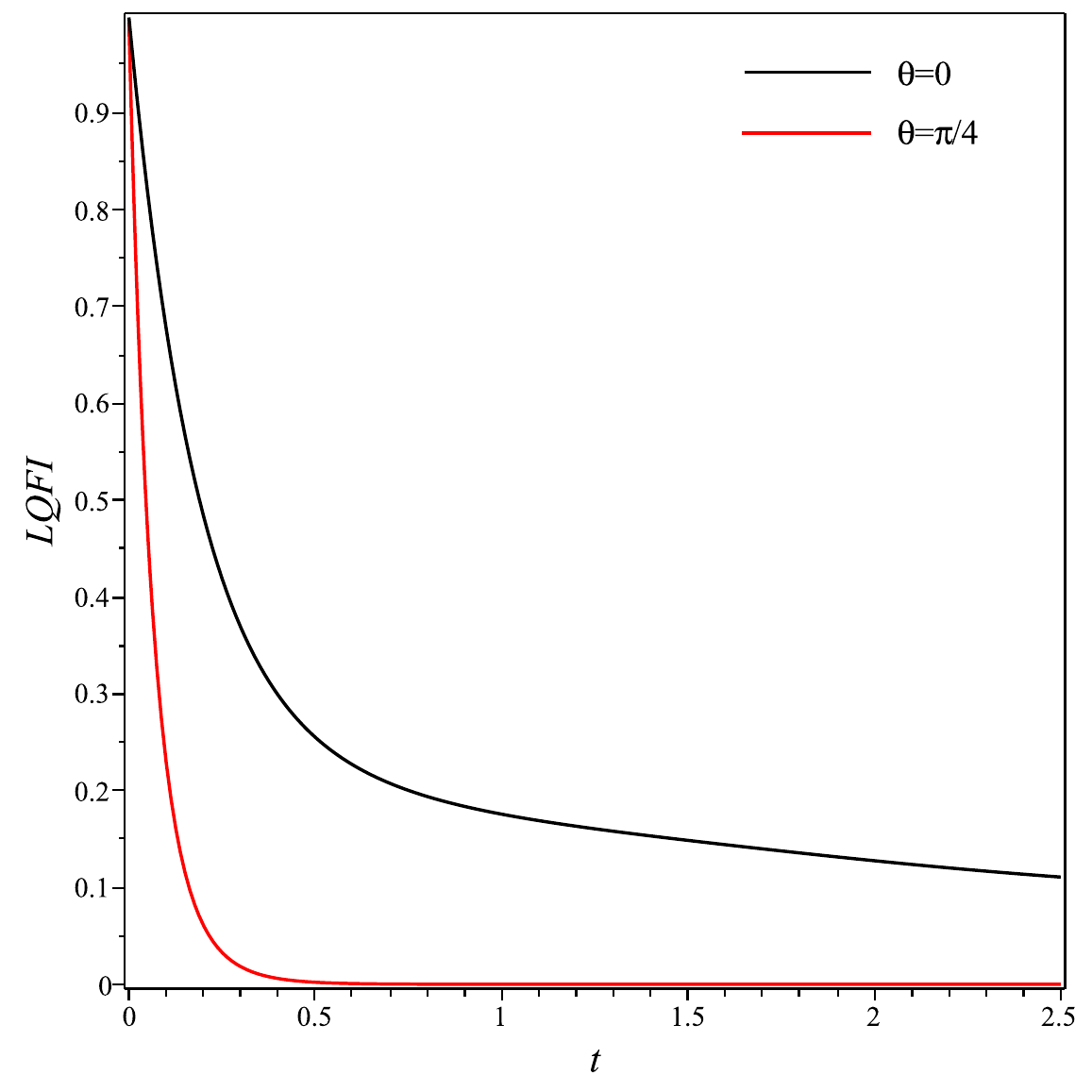}
    \caption{The local quantum Fisher information $(LQFI)$ as a function of time $t$ for different initial quantum states. The black curve represents the $LQFI$ for the non-entangled initial state $(\theta=0)$, while the red curve corresponds to the entangled initial state $(\theta=\pi/4)$. The parameters used are $\ell = 1$, $m=1$, $\epsilon=1$, and $\omega=0.5$.}
    \label{fig:LQFI}
\end{figure}
Fig.~\ref{fig:LQFI} illustrate the behavior of $LQFI$ as a function of time $t$, comparing the $LQFI$ for the initial state $(\theta=0)$ with that of the initial state $(\theta=\pi/4)$. This figure reveals that the quantum correlations embedded in the $LQFI$ for the initially non-entangled state are more robust than those for the initially entangled state. This occurs because entangled states are more sensitive to decoherence than separable states, causing the $LQFI$ to degrade more rapidly.

We summarize the qualitative behavior of the correlations in Table \ref{tab:summary}.

\begin{table}[h]
\centering
\begin{tabular}{|c|c|}
\hline
\textbf{Correlation} & \textbf{Effect of contact with quantum spacetime environment} \\
\hline
Concurrence (entanglement) & Destroyed in a finite time for $\theta\neq \pi/4$. Asymptotically vanishes otherwise. \\
\hline
$l_1$-norm of coherence & Asymptotically preserved for $\theta\neq0$. Asymptotically vanishes otherwise.  \\
\hline
Quantum Mutual Information & Asymptotically preserved for $\theta\neq0$. Asymptotically vanishes otherwise. \\
\hline
Local Quantum Fisher Information & Destroyed in finite time for $\theta=\pi/4$. Asymptotically vanishes or survives otherwise. \\
\hline
\end{tabular}
\caption{Qualitative summary of the behavior of the correlations.}
\label{tab:summary}
\end{table}


\section{The Role of the Different Planck Scale Contributions}\label{sec:diff_l}

To better understand the effect of the deformed Hamiltonian in creating correlations and the Lindbladians in destroying them, we introduce two deformation parameters, $\ell_H$ and $\ell_L$, to control these contributions. The Hamiltonian is given by:
\begin{align}
    {\cal H} &= \epsilon (\mathbb{1})^4 + \dfrac{\omega}{2} \left( \sigma_{z} \otimes \mathbb{1} \otimes \mathbb{1} \otimes \mathbb{1} + \mathbb{1} \otimes \sigma_{z} \otimes \mathbb{1} \otimes \mathbb{1} \right) \nonumber \\
    &\quad + \ell_H |P| \otimes |P| \otimes \sigma_{x} \otimes \sigma_{x},
\end{align}
and the Lindblad equation is given by:
\begin{equation}
    \partial_t \rho_{ab} + i \left[ {\cal H}_{ab}, \rho_{ab} \right] + \frac{\ell_L}{2} \left( {\cal P}_{ab}^2 \rho_{ab} + \rho_{ab} {\cal P}_{ab}^2 - 2 {\cal P}_{ab} \rho_{ab} {\cal P}_{ab} \right) = 0.
\end{equation}

\subsection{Lindbladian Weaker than Hamiltonian ($\ell_L \ll \ell_H$)}

\begin{figure}
    \centering
    \includegraphics[scale=0.6]{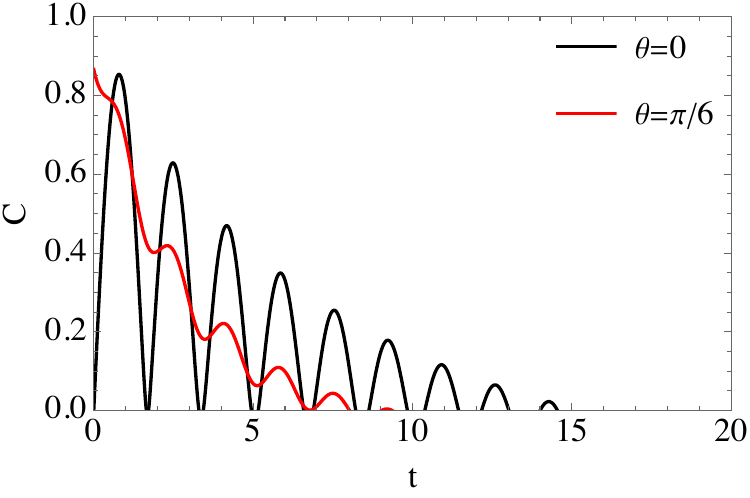}
    \caption{Concurrence for different initial conditions determined by $\theta$. The parameters used are $\ell_H = 1$, $\ell_L = 0.05$, $m = 1$, $\epsilon = 1$, and $\omega = 0.5$.}
    \label{fig:conc2}
\end{figure}

The effect of the different contributions to the Lindblad equation can be seen in Fig.~\ref{fig:conc2}, where we assume different values for the deformation parameters. The dissipation effect is described by $\ell_L = 0.05$, while the deformed Hamiltonian is described by $\ell_H = 1$. For an initially unentangled state ($\theta = 0$, black curve), we observe a dynamic phenomenon of alternating sudden death and sudden birth of quantum entanglement between the particles, whose amplitude is progressively attenuated by the dissipation introduced by the Lindblad term. In contrast, for a partially entangled initial state ($\theta = \pi/6$, blue curve), the concurrence drops quicker until it vanishes.

\begin{figure}
    \centering
    \includegraphics[scale=0.6]{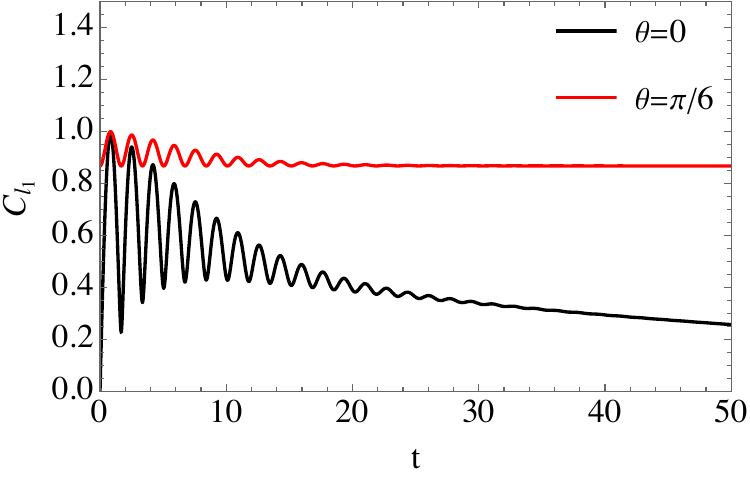}
    \caption{The $l_1$-norm of coherence for different initial conditions determined by $\theta$. The parameters used are $\ell_H = 1$, $\ell_L = 0.05$, $m = 1$, $\epsilon = 1$, and $\omega = 0.5$.}
    \label{fig:coh2}
\end{figure}

\begin{figure}
    \centering
    \includegraphics[scale=0.6]{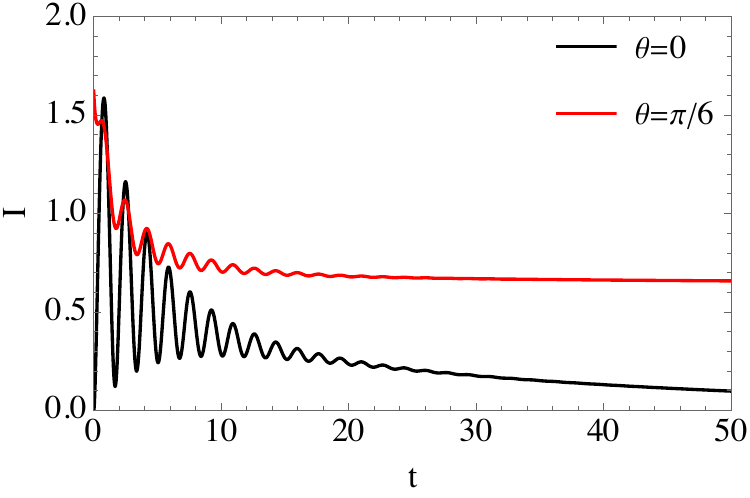}
    \caption{Quantum mutual information for different initial conditions determined by $\theta$. The parameters used are $\ell_H = 1$, $\ell_L = 0.05$, $m = 1$, $\epsilon = 1$, and $\omega = 0.5$.}
    \label{fig:disc2}
\end{figure}

Similarly, the $l_1$-norm of coherence and quantum mutual information exhibit oscillatory behavior, as shown in Figs.~\ref{fig:coh2} and \ref{fig:disc2}. For an initially unentangled state ($\theta = 0$, black curve) and for the partially entangled state ($\theta = \pi/6$, red curve), the coherence oscillates but tends to stabilize over time, as observed in the case where $\ell_H = \ell_L$. To further study the role of these contributions, let us consider the opposite case.

\subsection{Hamiltonian Weaker than Lindbladian ($\ell_H \ll \ell_L$)}

In this case, when the system starts in an unentangled state, the deformed composition law contribution is not strong enough to produce non-zero concurrence. For an initially entangled state, the concurrence evolves similarly to the behavior shown in Fig.~\ref{fig:conc} for the same set of parameters.

Regarding the $l_1$-norm of coherence and quantum mutual information, the system behaves similarly to Figs.~\ref{fig:coh} and \ref{fig:qdisc}. This provides evidence that the quantum spacetime, which acts as an environment of the two-qubit system, actively preserving the coherence of the system. Although it dissipates entanglement, it helps create and preserve classical and quantum correlations beyond entanglement. The reason for this behavior is in the fact that the Lindbladian $P_{ab}$ is not diagonal due to the superposition of momenta in this model, as can be seen in \eqref{eq:pab}. This property is responsible for giving a non-null contribution from the open system represented by the quantum spacetime and creating superpositions in this model.


\section{Discussion}\label{sec:discussion}

We analyzed how modifications to the Hamiltonian of a system of particles, arising from deformations of the relativistic symmetries, influence quantum correlations. Specifically, we constructed a model describing two particles in a superposition of energies and momenta, corresponding to a system of four qubits. This system exhibits intriguing properties, such as the emergence of entanglement due to the modified Hamiltonian and the role of the quantum spacetime as an environment that dissipates entanglement while simultaneously creating and preserving other classical and quantum correlations between the particles. This behavior was observed when we individually examined the contributions of the quantum gravity parameters from the deformed energy composition law (Hamiltonian) and from the Lindblad equation.

The unique ability of the deformed Hamiltonian to generate entanglement between the particles in the system is a distinctive effect of a deformation of relativistic symmetries. 
If coherence is maintained during the time scale of the entanglement time described in this paper, then systems of qubits could be considered to test deformations of the composition of energy due to quantum gravity. The study of $N$-partite systems could also be beneficial to understand the effect of quantum gravity on more general systems \cite{Wu:2021pja}.


\appendix

\section{Analytic solution of the Lindblad equation and asymptotic behavior of the density matrix}\label{sec:app1}

In this section, we drop the notation with an overbar for the density matrix for simplicity. The Lindblad equation can be cast in this form
\[
\begin{pmatrix}
A_{11} & 0 & 0 & A_{14} \\
0 & A_{22} & A_{23} & 0 \\
0 & A_{32} & A_{33} & 0 \\
A_{41} & 0 & 0 & A_{44}
\end{pmatrix}=0,
\]

where

\begin{align}
&A_{11} =(2-2 i) \alpha  \ell m \rho _{14}-2 \alpha  \ell m \rho _{23}-2 \alpha  \ell m \rho _{32}+(2+2 i) \alpha  \ell m \rho _{41}+8 \ell m \epsilon  \rho _{11}-4 \ell m \epsilon  \rho _{22}-4 \ell m \epsilon  \rho _{33}+4 \rho _{11}’,\\
&A_{44} =(2+2 i) \alpha  \ell m \rho _{14}-2 \left(\alpha  \ell m \rho _{23}+\alpha  \ell m \rho _{32}-(1-i) \alpha  \ell m \rho _{41}+2 \ell m \epsilon  \rho _{22}+2 \ell m \epsilon  \rho _{33}-4 \ell m \epsilon  \rho _{44}-2 \rho _{44}'\right),\\
&A_{14} =(2-2 i) \alpha  \ell m \rho _{11}-2 \alpha  \ell m \rho _{22}-2 \alpha  \ell m \rho _{33}+(2+2 i) \alpha  \ell m \rho _{44}+8 \ell m \epsilon  \rho _{14}-4 \ell m \epsilon  \rho _{23}-4 \ell m \epsilon  \rho _{32}+4 \rho _{14}’(t)\\
&A_{41} =(2+2 i) \alpha  \ell m \rho _{11}-2 \left(\alpha  \ell m \rho _{22}+\alpha  \ell m \rho _{33}-(1-i) \alpha  \ell m \rho _{44}+2 \ell m \epsilon  \rho _{23}+2 \ell m \epsilon  \rho _{32}-4 \ell m \epsilon  \rho _{41}-2 \rho _{41}'\right),\\
&A_{22} =-2 \alpha  \ell m \rho _{14}+(2-2 i) \alpha  \ell m \rho _{23}+(2+2 i) \alpha  \ell m \rho _{32}-2 \alpha  \ell m \rho _{41}-4 \ell m \epsilon  \rho _{11}+8 \ell m \epsilon  \rho _{22}-4 \ell m \epsilon  \rho _{44}+4 \rho _{22}’(t),\\
&A_{23} =-2 \alpha  \ell m \rho _{11}+(2-2 i) \alpha  \ell m \rho _{22}+(2+2 i) \alpha  \ell m \rho _{33}-2 \alpha  \ell m \rho _{44}-4 \ell m \epsilon  \rho _{14}+8 \ell m \epsilon  \rho _{23}-4 \ell m \epsilon  \rho _{41}+4 \rho _{23}’(t),\\
&A_{32} =-2 \alpha  \ell m \rho _{11}+(2+2 i) \alpha  \ell m \rho _{22}+(2-2 i) \alpha  \ell m \rho _{33}-2 \alpha  \ell m \rho _{44}-4 \ell m \epsilon  \rho _{14}+8 \ell m \epsilon  \rho _{32}-4 \ell m \epsilon  \rho _{41}+4 \rho _{32}’(t),\\
&A_{33} =-2 \alpha  \ell m \rho _{14}+(2+2 i) \alpha  \ell m \rho _{23}+(2-2 i) \alpha  \ell m \rho _{32}-2 \alpha  \ell m \rho _{41}-4 \ell m \epsilon  \rho _{11}+8 \ell m \epsilon  \rho _{33}-4 \ell m \epsilon  \rho _{44}+4 \rho _{33}',
\end{align}
and $\alpha=\sqrt{\epsilon^2-\omega^2}+\epsilon$.

Subtracting the equations for $A_{11}$ and $A_{44}$, we find

\begin{equation}
    -i \alpha  \ell m \rho _{14}(t)+i \alpha  \ell m \rho _{41}(t)+2 \ell m \epsilon  \rho _{11}(t)-2 \ell m \epsilon  \rho _{44}(t)+\rho _{11}'(t)-\rho _{44}'(t)=0.
\end{equation}

Subtracting the equations for $A_{14}$ and $A_{41}$, we find
\begin{equation}
    -i \alpha  \ell m \rho _{11}(t)+i \alpha  \ell m \rho _{44}(t)+2 \ell m \epsilon  \rho _{14}(t)-2 \ell m \epsilon  \rho _{41}(t)+\rho _{14}'(t)-\rho _{41}'(t)=0.
\end{equation}

Let us call $\rho_{11}-\rho_{44}=A$ and $\rho_{14}-\rho_{41}=B$, where $B$ is two times the imaginary part of $\rho_{14}$. We then have
\begin{align}
   & -i \alpha  \ell m B(t)+2 \ell m \epsilon  A(t)+A'(t)=0,\\
    &-i \alpha  \ell m A(t)+2 \ell m \epsilon  B(t)+B'(t)=0.
\end{align}

With initial conditions $A(0)=0=B(0)$ (according to \eqref{eq:init_cond2}), the solution is $A(t)=0=B(t)$. This proves that $\rho_{11}=\rho_{44}$ and that $\rho_{14}=\rho_{41}=\text{Re}(\rho_{14})$ is a real function. We shall call $\rho_{11}(t)=a(t)$ and $\rho_{14}(t)=b(t)$.

Subtracting the equations for $A_{22}$ and $A_{33}$, we find

\begin{equation}
    -i \alpha  \ell m \rho _{23}(t)+i \alpha  \ell m \rho _{32}(t)+2 \ell m \epsilon  \rho _{22}(t)-2 \ell m \epsilon  \rho _{33}(t)+\rho _{22}'(t)-\rho _{33}'(t)=0.
\end{equation}

Subtracting the equations for $A_{23}$ and $A_{32}$, we find
\begin{equation}
    -i \alpha  \ell m \rho _{22}(t)+i \alpha  \ell m \rho _{33}(t)+2 \ell m \epsilon  \rho _{23}(t)-2 \ell m \epsilon  \rho _{32}(t)+\rho _{23}'(t)-\rho _{32}'(t)=0.
\end{equation}

Let us call $\rho_{22}-\rho_{33}=C$ and $\rho_{23}-\rho_{32}=D$, where $D$ is two times the imaginary part of $\rho_{23}$. We then have
\begin{align}
   & -i \alpha  \ell m D(t)+2 \ell m \epsilon  C(t)+C'(t)=0,\\
    &-i \alpha  \ell m C(t)+2 \ell m \epsilon  D(t)+D'(t)=0.
\end{align}

With initial conditions $C(0)=\cos^2{\theta}-\sin^{2}(\theta)=\cos(2\theta)$ (according to \eqref{eq:init_cond2}) and $D(0)=0$, the solution is 
\begin{align}
    &C(t)=\rho_{22}(t)-\rho_{33}(t)=\cos (2 \theta ) e^{-2 \ell m t \epsilon } \cos (\alpha  \ell m t),\label{eq:C}\\
    &D(t)=2\text{Im}(\rho_{23}(t))=\rho_{23}(t)-\rho_{32}(t)=i \cos (2 \theta ) e^{-2 \ell m t \epsilon } \sin (\alpha  \ell m t).
\end{align}

Now, let us sum up the terms from $A_{23}$ and $A_{32}$:
\begin{align}
    -2 \alpha  \ell m a(t)-4 \ell m \epsilon  b(t)+\alpha  \ell m \rho _{22}(t)+\alpha  \ell m \rho _{33}(t)+2 \ell m \epsilon  \rho _{23}(t)+2 \ell m \epsilon  \rho _{32}(t)+\rho _{23}'(t)+\rho _{32}'(t)=0.
\end{align}

Let us call $\rho_{23}(t)+\rho_{32}(t)=F(t)$. Also, we notice that due to the property $\text{Tr}(\rho)=1$, we have $\rho_{22}+\rho_{33}=1-2a(t)$. This means that we can simplify the above equation to
\begin{equation}
    F'(t)-4 \alpha  \ell m a(t)-4 \ell m \epsilon  b(t)+2 \ell m \epsilon  F(t)+\alpha  \ell m=0.\label{eq:F}
\end{equation}

Also, the equations for $A_{11}$ and $A_{14}$ can be written as
\begin{align}
    &a'(t)+4 \ell m \epsilon  a(t)+\alpha  \ell m b(t)-\frac{1}{2} \alpha  \ell m F(t)-l m \epsilon=0,\label{eq:a}\\
   & b'(t)+2 \alpha  \ell m a(t)+2 \ell m \epsilon  b(t)-l m \epsilon  F(t)-\frac{1}{2} \alpha  \ell m=0.\label{eq:b}
\end{align}

The solutions of equations \eqref{eq:a}, \eqref{eq:b} and \eqref{eq:F}, for the initial conditions $a(0)=0=b(0)$ and $F(0)=\sin(2\theta)$, are
\begin{align}
    &a(t)=\rho_{11}(t)=\rho_{44}(t)=\frac{1}{4}-\frac{1}{8} e^{-2 \ell m t (\sqrt{\epsilon ^2-\omega ^2}+3 \epsilon)}(1+\sin(2\theta))-\frac{1}{8}e^{-2 \ell m t(\epsilon-\sqrt{\epsilon^2-\omega^2})}(1-\sin(2\theta)),\\
    &b(t)=\rho_{14}(t)=\frac{1}{4} \sin (2 \theta )-\frac{1}{8} e^{-2 \ell m t (\sqrt{\epsilon ^2-\omega ^2}+3 \epsilon)}(1+\sin(2\theta))+\frac{1}{8}e^{-2 \ell m t(\epsilon-\sqrt{\epsilon^2-\omega^2})}(1-\sin(2\theta)),\\
    &F(t)=2\text{Re}(\rho_{23}(t))=\frac{1}{2} \sin (2 \theta )+\frac{1}{4} e^{-2 \ell m t (\sqrt{\epsilon ^2-\omega ^2}+3 \epsilon)}(1+\sin(2\theta))-\frac{1}{4}e^{-2 \ell m t(\epsilon-\sqrt{\epsilon^2-\omega^2})}(1-\sin(2\theta)),
\end{align}
where we used the definition of $\alpha=\sqrt{\epsilon^2-\omega^2}+\epsilon$.

This implies that
\begin{equation}
    \rho_{23}(t)=\frac{1}{4} \sin (2 \theta )+\frac{1}{8} e^{-2 \ell m t (\sqrt{\epsilon ^2-\omega ^2}+3 \epsilon)}(1+\sin(2\theta))-\frac{1}{8}e^{-2 \ell m t(\epsilon-\sqrt{\epsilon^2-\omega^2})}(1-\sin(2\theta))+\frac{i}{2} \cos (2 \theta ) e^{-2 \ell m t \epsilon } \sin (\alpha  \ell m t).
\end{equation}

From \eqref{eq:C} and the fact that $\rho_{22}+\rho_{33}=1-2a(t)$, we derive the following solutions
\begin{align}
    &\rho _{22}(t)=\frac{1}{8}\left(2+4 \cos (\theta ) e^{-2 \ell m t \epsilon } \cos (\alpha  \ell m t)+e^{-2 \ell m t(\epsilon-\sqrt{\epsilon^2-\omega^2})}(e^{-4 \ell m t (\sqrt{\epsilon ^2-\omega ^2}+\epsilon)}-(\sin (2 \theta )-1)+\sin (2 \theta ))\right),\\
    &\rho_{33}(t)=\frac{1}{8}\left(2-4 \cos (\theta ) e^{-2 \ell m t \epsilon } \cos (\alpha  \ell m t)+e^{-2 \ell m t(\epsilon-\sqrt{\epsilon^2-\omega^2})}(e^{-4 \ell m t (\sqrt{\epsilon ^2-\omega ^2}+\epsilon)}-(\sin (2 \theta )-1)+\sin (2 \theta ))\right).
\end{align}

Therefore, we have the complete set of solutions of the Lindblad equation for our problem. From these expressions, we can derive the asymptotic behavior of the density matrix when $t\rightarrow \infty$:
\begin{align}
&\lim_{t\rightarrow \infty}\rho_{11}=\lim_{t\rightarrow \infty}\rho_{22}=\lim_{t\rightarrow \infty}\rho_{33}=\lim_{t\rightarrow \infty}\rho_{44}=\frac{1}{4},\\
    &\lim_{t\rightarrow \infty}\rho_{14}=\lim_{t\rightarrow \infty}\rho_{23}=\frac{\sin({\theta})\cos(\theta)}{2}=\frac{\sin{(2\theta)}}{4}.\label{eq:assymp_appA}
\end{align}

This means that asymptotically, the off-diagonal elements of the density matrix are non-zero for an initially entangled (or partially entangled) state. It also means that the diagonal elements of the density matrix become larger than the off-diagonal ones. For an initially entangled state ($\theta=\pi/4$), they become equal at asymptotic times.

\section{Quantum Mutual Information of an $X$-state}\label{sec:app2}

Following the steps described in Section \ref{sec:disc}, the quantum mutual information for a bipartite system of $X$-shape (represented by a $4 \times 4$ density matrix):
\begin{equation}
\begin{aligned}
I=\frac{1}{2 \ln 2} \Bigg[
    & -2 (\rho_{11} + \rho_{22}) \ln(\rho_{11} + \rho_{22}) -2 (\rho_{11} + \rho_{33}) \ln(\rho_{11} + \rho_{33}) \\
    & + \left( \rho_{22} + \rho_{33} - \sqrt{ (\rho_{22} - \rho_{33} )^2+4 |\rho_{23}|^2 } \right)
        \ln \left( \frac{1}{2} \left( \rho_{22} + \rho_{33} - \sqrt{(\rho_{22} - \rho_{33} )^2+4 |\rho_{23}|^2 } \right) \right) \\
    & + \left( \rho_{22} + \rho_{33} + \sqrt{ (\rho_{22} - \rho_{33} )^2+4 |\rho_{23}|^2 } \right)
        \ln \left( \frac{1}{2} \left( \rho_{22} + \rho_{33} + \sqrt{ (\rho_{22} - \rho_{33} )^2+4 |\rho_{23}|^2 } \right) \right) \\
    & -2 (\rho_{22} + \rho_{44}) \ln(\rho_{22} + \rho_{44}) -2 (\rho_{33} + \rho_{44}) \ln(\rho_{33} + \rho_{44}) \\
    & + \left( \rho_{11} + \rho_{44} - \sqrt{ (\rho_{11}-\rho_{44})^2 + 4 |\rho_{14}|^2 } \right)
        \ln \left( \frac{1}{2} \left( \rho_{11} + \rho_{44} - \sqrt{ (\rho_{11}-\rho_{44})^2 + 4 |\rho_{14}|^2 } \right) \right) \\
    & + \left( \rho_{11} + \rho_{44} + \sqrt{ (\rho_{11}-\rho_{44})^2 + 4 |\rho_{14}|^2 } \right)
        \ln \left( \frac{1}{2} \left( \rho_{11} + \rho_{44} + \sqrt{ (\rho_{11}-\rho_{44})^2 + 4 |\rho_{14}|^2 } \right) \right)
\Bigg].
\end{aligned}
\end{equation}

\section*{Acknowledgments}
I. P. L. would like to thank Adriano Braga Barreto for the insightful discussions.
I. P. L. was partially supported by the National Council for Scientific and Technological Development - CNPq,
grant 312547/2023-4. G. V. thanks Coordena\c c\~ao de Aperfei\c coamento de Pessoal de Nível Superior - Brazil (CAPES) - Finance Code 001 and the National Council for Scientific and Technological Development - CNPq,
grant 140335/2022-6 for financial support. 
V. B. B. was partially supported by the National Council for Scientific and Technological Development- CNPq, grant No. 307211/2020 7.  M. R.
was partially supported by the National Council for Scientific and Technological Development - CNPq,
grant 311565/2025-5. G.G. is grateful for financial support by the Programme STAR Plus, funded by Federico II University and Compagnia di San Paolo, and by the MIUR, PRIN 2017 grant 20179ZF5K.

The authors would like to acknowledge networking support by the COST Action BridgeQG (CA23130) and the COST Action RQI (CA23115), supported by COST (European Cooperation in Science and Technology).
\bibliographystyle{utphys}
\bibliography{dsr-ent}
\end{document}